\pdfoutput=1
\documentclass{article}
\usepackage{iclr2027_conference,times}
\setcitestyle{numbers,square,comma}
\usepackage{graphicx}
\usepackage{tikz}
\usetikzlibrary{positioning,fit,backgrounds,calc,arrows.meta}
\usepackage{xcolor,array,multirow,booktabs,tabularx,placeins,amsmath}
\usepackage{hyperref,url}
\newcommand{\Description}[1]{}
\newcommand{\rulesource}[1]{\par{\footnotesize\citep{#1}}}

\title{Exploring Causal Mechanisms with Generative Agent-Based Models}
\author{Xuan Liu$^{1}$, Haoyang Shang$^{2}$, Tanya Bhat$^{1}$, Haojian Jin$^{1}$ \\
$^{1}$University of California, San Diego \\
$^{2}$University of British Columbia \\
\texttt{xul049@ucsd.edu}}
\iclrfinalcopy
\begin{document}
\maketitle
\begin{abstract}


In this paper, we explore using generative agent-based models for a classical ABM application: testing how individual-level behavioral rules produce collective phenomena. We introduce RePair, a method that calibrates simulation worlds, operationalizes candidate mechanisms as natural-language rules, estimates their effects through matched interventions, and examines behavioral traces. We assess the method by testing it in four simulation worlds grounded in established social-science models and empirical studies. Our results reveal that (1) natural-language rules can produce measurable collective effects; (2) rule comparisons can converge as configurations accumulate; and (3) behavioral traces connect collective effects to agents' actions and interactions, helping researchers evaluate the proposed causal process. Together, these findings show the feasibility of using generative agent-based models to explore causal mechanisms and provide practical guidance for producing reliable, interpretable explanations.

\end{abstract}



\section{Introduction}
\label{sec:intro}

Agent-based modeling (ABM) is widely used to explore how individual-level behavioral rules produce collective phenomena such as segregation, cooperation, and polarization~\citep{anderson1972more,epstein1999agent,macy2002factors,bruch2015empirical}. Researchers encode a hypothesized mechanism as a local rule, intervene on that rule, and observe its consequences for a collective outcome. Schelling's segregation model provides a canonical example: households relocate when too few neighbors belong to their own group, and these local decisions can transform an initially mixed neighborhood into a segregated one~\citep{schelling1971dynamic}. Such experiments do not by themselves establish how people behave in the world, but they can show whether a proposed mechanism is sufficient to generate, amplify, or suppress a phenomenon within a specified model.

Generative agent-based models extend this application by using large language model (LLM) agents to simulate individual decisions and social interactions~\citep{aher2023simulating,mei2024turing}. Recent work has used LLM agents to simulate town life~\citep{park2023generative}, resource dilemmas~\citep{piatti2024cooperate}, peer review~\citep{jin2024agentreview}, opinion dynamics~\citep{chuang2024opinion}, and open-ended social interaction~\citep{liu2024sociallyaligned}. These simulations can produce recognizable collective patterns, including cooperation~\citep{piatti2024cooperate}, shared conventions, and collective bias~\citep{ashery2025conventions}. However, existing research primarily demonstrates that these patterns can emerge; it does not systematically test whether introducing a specific behavioral rule produces a predicted collective outcome. Because different rules can generate the same aggregate pattern~\citep{epstein2023inverse,leonmedina2017sufficiency}, observing the pattern alone cannot identify its causal mechanism~\citep{larooij2025large,zhao2026mechanism}. Researchers therefore still lack a method for intervening on candidate rules, measuring their collective effects, and assessing whether behavioral evidence supports the proposed causal process.

\begin{figure*}[t]
  \centering
  \includegraphics[width=\textwidth]{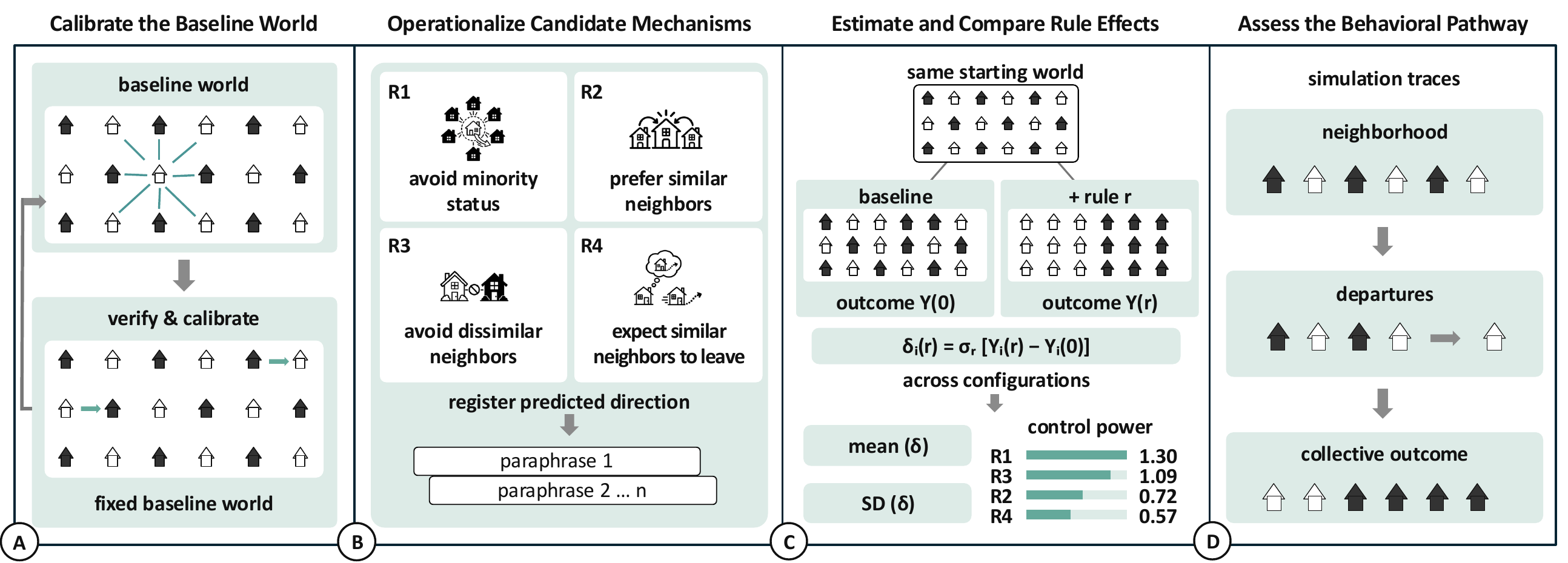}
  \caption{\textbf{Overview of RePair's four-step method}, illustrated with residential segregation. Panels A--D correspond to Steps 1--4 in Section~\ref{sec:protocol}.
  \textbf{(A)} Calibrate and fix a baseline world that executes correctly, supports the relevant actions and interactions, and leaves room for the outcome to change.
  \textbf{(B)} Operationalize candidate mechanisms as local rules, register their predicted effects, and prepare semantically equivalent wordings.
  \textbf{(C)} Compare each rule condition with a matched baseline and aggregate signed outcome differences across configurations to estimate effects, quantify uncertainty, and compare rules.
  \textbf{(D)} Examine behavioral traces to assess whether the observed actions and their temporal sequence support the proposed pathway to the collective outcome.}
  \Description{Four panels illustrate the method with residential segregation. Panel A shows a loop of baseline verification and calibration that yields a fixed baseline world. Panel B presents four candidate household rules, registers their predicted directions, and prepares alternative wordings. Panel C compares each rule condition with a matched baseline from the same starting world, calculates signed outcome differences, and summarizes their mean and standard deviation across configurations to compare rules by control power. Panel D examines simulation traces, showing the neighborhood, household departures, and the collective outcome to assess the proposed behavioral pathway.}
  \label{fig:protocol-overview}
\end{figure*}

We introduce RePair, a method for exploring causal mechanisms with generative agent-based models. We define a candidate mechanism as an account of how an individual behavior, through interaction, produces a collective outcome. The method evaluates this account in four steps (Figure~\ref{fig:protocol-overview}). First, it calibrates a baseline simulation world by verifying that the world executes as intended, permits the actions and interactions required by the mechanism, and gives the outcome room to change. Second, it operationalizes the mechanism as a natural-language local rule with a registered prediction about its effect on a collective outcome. Third, it compares each rule condition with a matched baseline that holds the model, agent profiles, world description, initial conditions, and interaction schedule fixed. The paired outcome difference estimates the effect of adding the rule within that simulated world. Repeating this comparison across configurations estimates the mean effect and its uncertainty. We also report \emph{control power}, which relates the effect's magnitude to its variation across configurations and supports comparisons among rules. Tests across models, semantically equivalent rule wordings, and sentence positions determine whether a conclusion extends beyond one implementation. Finally, behavioral traces show whether the expected actions occurred and whether the observed interactions and temporal sequence support the proposed causal process. The method thus separates three claims that an aggregate pattern alone conflates: the rule changed agent behavior, the collective outcome changed, and the behavioral evidence supports the proposed link between them.

We assess the feasibility of this method through three research questions using four simulation worlds grounded in established social-science models and experiments: residential segregation~\citep{schelling1971dynamic,card2008tipping}, public-goods cooperation~\citep{fehr2000cooperation}, group polarization~\citep{moscovici1969group}, and cooperative cascades~\citep{fowler2010cooperative}.

\textbf{RQ1: Can hypothesized individual-level mechanisms be implemented as interventions whose effects on collective outcomes can be measured and compared?} We calibrate each world, introduce candidate mechanisms as natural-language rules, and compare every rule condition with its matched baseline. The residential-segregation study demonstrates the complete process. We first establish a world in which households can act on relocation preferences and the departure-rate gap between groups can increase or decrease. We then compare four rules predicted to widen that gap and four predicted to narrow it. The resulting effects distinguish mechanisms that change the same outcome in different directions and by different amounts: for example, minority-status aversion produces the most consistent amplifying effect, whereas an integration preference produces the clearest suppressive effect. Across all four worlds, each world contains at least one rule with an established effect in its predicted direction; among all 23 candidate rules, 12 have 90\% intervals entirely in that direction. These results show that natural-language rules can serve as measurable interventions when the calibrated world supports the relevant behavior and a responsive outcome.

\textbf{RQ2: Do effect estimates and rule comparisons converge across configuration samples, models, rule wordings, and sentence positions?} We study this question primarily in the public-goods world using 268 configurations across four LLMs and five candidate rules. Model-balanced resampling shows how conclusions change as configurations accumulate. The leading rule is recovered in every resample from eight configurations onward, and the ordering of the ten rule pairs distinguished by the full sample is preserved in 99.98\% of resamples at 268. Recovering every rule's effect conclusion is harder, reaching 80.4\% agreement at 40 configurations and 94.2\% at 268, because rules near zero remain uncertain. Reliability also depends on the claim being made. Some effects persist across models and alternative formulations, but inequity-aversion wordings can move the outcome in opposite directions, and moving the same sentence within the profile can change its effect. Additional samples improve precision; variation across models, wordings, and sentence positions instead establishes the range of implementations to which a conclusion applies.

\textbf{RQ3: Do behavioral traces connect measured collective effects to the actions and interactions proposed by a candidate mechanism?} We inspect whether agents perform the behavior named by a rule, whether that behavior propagates through interaction, and whether it precedes the collective change. The segregation traces, for example, separate changes in the departure-rate gap from changes in each group's relocation rate and timing, revealing how rules with similar aggregate effects operate differently. The group-polarization and cooperative-cascade studies show why this evidence matters. In the polarization world, rules can increase discussion without producing more extreme group opinions. In the cascade world, contribution effects can appear in the first period, before agents observe their partners and before behavioral contagion could operate. Thus, an intervention may change the registered outcome without supporting the proposed explanation. Behavioral traces identify which link is supported---the response to the rule, its propagation through interaction, or the aggregate outcome---and which link requires a follow-up intervention. Together, the three evaluations demonstrate that generative ABMs can support causal exploration when effect estimates and process evidence are interpreted jointly, with claims bounded to the tested simulation worlds and configurations.

This work makes three contributions. First, we formulate generative ABM as an instrument for testing how hypothesized individual-level behaviors produce collective phenomena, with causal claims bounded to the specified simulation. Second, we contribute a method that integrates world calibration, matched interventions, reliability assessment, and behavioral-pathway analysis. Third, we provide empirical evidence across four collective phenomena and practical guidance for determining when candidate mechanisms can be measured, compared, and interpreted. Generative ABMs can help researchers screen and refine causal explanations before evaluating them against human experiments or observational data; they do not replace that external empirical validation.

\section{Related Work}
We review how agent-based models explain collective emergence, how LLM-agent simulations reproduce social patterns, and how interactive simulations help people understand complex systems.

\subsection{Emergence and Generative Explanation in Agent-Based Models}

Agent-based models (ABMs) explain collective phenomena by specifying how
individual agents respond to local conditions and examining the macro patterns
produced through repeated interaction
\citep{macy2002factors,bruch2015empirical,bonabeau2002agent}. Classic models show how local
attitudes can generate spatial structures \citep{sakoda1971checkerboard}, mild
neighborhood preferences can produce segregation
\citep{schelling1971dynamic}, individual thresholds can yield sharply different
levels of collective participation \citep{granovetter1978threshold}, and local
convergence can coexist with global polarization
\citep{axelrod1997dissemination}, among other collective phenomena. Across these
models, the macro pattern is not specified directly; it emerges from executable
local rules~\citep{anderson1972more,bonabeau2002agent}.

This bottom-up relationship supports a generative account of explanation:
researchers explain a social regularity by showing how it can be ``grown'' from
specified agents and interactions
\citep{epstein1999agent,hedstrom2010mechanisms}. Yet generating a target pattern
does not establish that its mechanism is unique or empirically valid. Different
rules and parameterizations may produce similar outcomes
\citep{epstein2023inverse,leonmedina2017sufficiency}, motivating comparison among
candidate mechanisms, sensitivity analysis, and explicit model documentation
\citep{grimm2005pattern,windrum2007empirical,grimm2006odd}.

Generative ABMs retain this explanatory aim but change the status of the local
rule: an LLM interprets a natural-language description rather than executing a
fixed policy. Our work asks how ABM's experimental use can be retained under
this change. We retain the practices of comparing candidates, testing
sensitivity, and registering the design, while explicitly estimating the
effects and uncertainty of rules whose implementation depends on LLM
interpretation.

\subsection{Emergent Social Behavior in LLM-Agent Simulations}

LLM-agent simulations extend social simulation with agents that communicate,
remember interactions, and generate context-sensitive behavior in natural
language \citep{anthis2025position,park2023generative}. Recent work has used such populations to
simulate social computing communities, everyday social interaction, opinion
dynamics, cooperation, social alignment, norm formation, conventions, and
institutional processes
\citep{park2022socialsimulacra,park2023generative,liu2024sociallyaligned,
zhou2024sotopia,chuang2024opinion,piatti2024cooperate,ren2024norms,
jin2024agentreview,ashery2025conventions,liu2025exploring,shang2026united}. Collectively, these studies show that
repeated interactions among LLM agents can produce recognizable macro-level
patterns, while also revealing substantial dependence on model choice,
communication, information structure, prompting, and simulation design
\citep{chuang2024opinion,piatti2024cooperate,zhou2024misleading,ashkinaze2025plurals,liu2026cobra}.

This literature establishes that collective patterns can arise in generative
ABMs, but the appearance of a pattern does not identify which candidate rule
produced it~\citep{zhao2026mechanism}. Our work shifts the basis of evidence from the generated pattern alone to the measured effect of a corresponding intervention. It then asks whether the estimated effect is stable across configurations, including different models and wordings, given the sensitivities just noted. It also asks whether behavioral traces support the actions, interactions, and temporal order proposed by the candidate mechanism.

\subsection{Interactive Tools for Systematic Experimentation}

HCI and learning-sciences research has long used interactive simulation to help
people reason about complex systems
\citep{papert1980mindstorms,resnick1994turtles,resnick1998diving}. In
traditional agent-based environments, agents follow executable, rule-based
policies; users can modify these local rules, run the model, and inspect the
system-level patterns that follow
\citep{repenning1993agentsheets,wilensky1999participatory,tisue2004netlogo,smith1994kidsim}.
By allowing users to manipulate local rules and observe the resulting macro
outcomes, these environments help users understand how coordinated
system-level patterns can arise from repeated local interactions without
centralized control
\citep{resnick1998diving,wilensky1999levels,wilensky1999participatory}.

Recent HCI research has extended this experimental orientation to generative
AI by supporting parallel variations of prompts and model configurations,
side-by-side comparison of outputs, user-defined evaluation criteria, and
inspection of intermediate results
\citep{wu2022aichains,kim2023llmobjects,arawjo2024chainforge,kim2024evallm,
kahng2024llmcomparator,shankar2024validators}. These approaches help users
investigate model behavior through structured comparison and repeated
evaluation, rather than drawing conclusions from a single plausible output.

These systems make generative-model behavior easier to compare and inspect. In a generative ABM, however, the object of study is not a single model response: an intervention shapes the decisions of many agents, those decisions unfold through repeated interactions, and evidence appears at both the behavioral and collective levels. Our work provides an experimental method for this micro-to-macro setting. 




\section{RePair: A Method for Exploring Causal Mechanisms}
\label{sec:protocol}


Figure~\ref{fig:protocol-overview} illustrates an overview of our method, which contains four steps. Below, we first describe how we build the agent-based simulation environment and then describe the steps that measure the effect of an individual mechanism on a collective outcome. 

\subsection{Building the Simulation Environment}
\label{sec:protocol-terms}

\begin{figure}[!htbp]
    \centering
    \includegraphics[width=\textwidth]{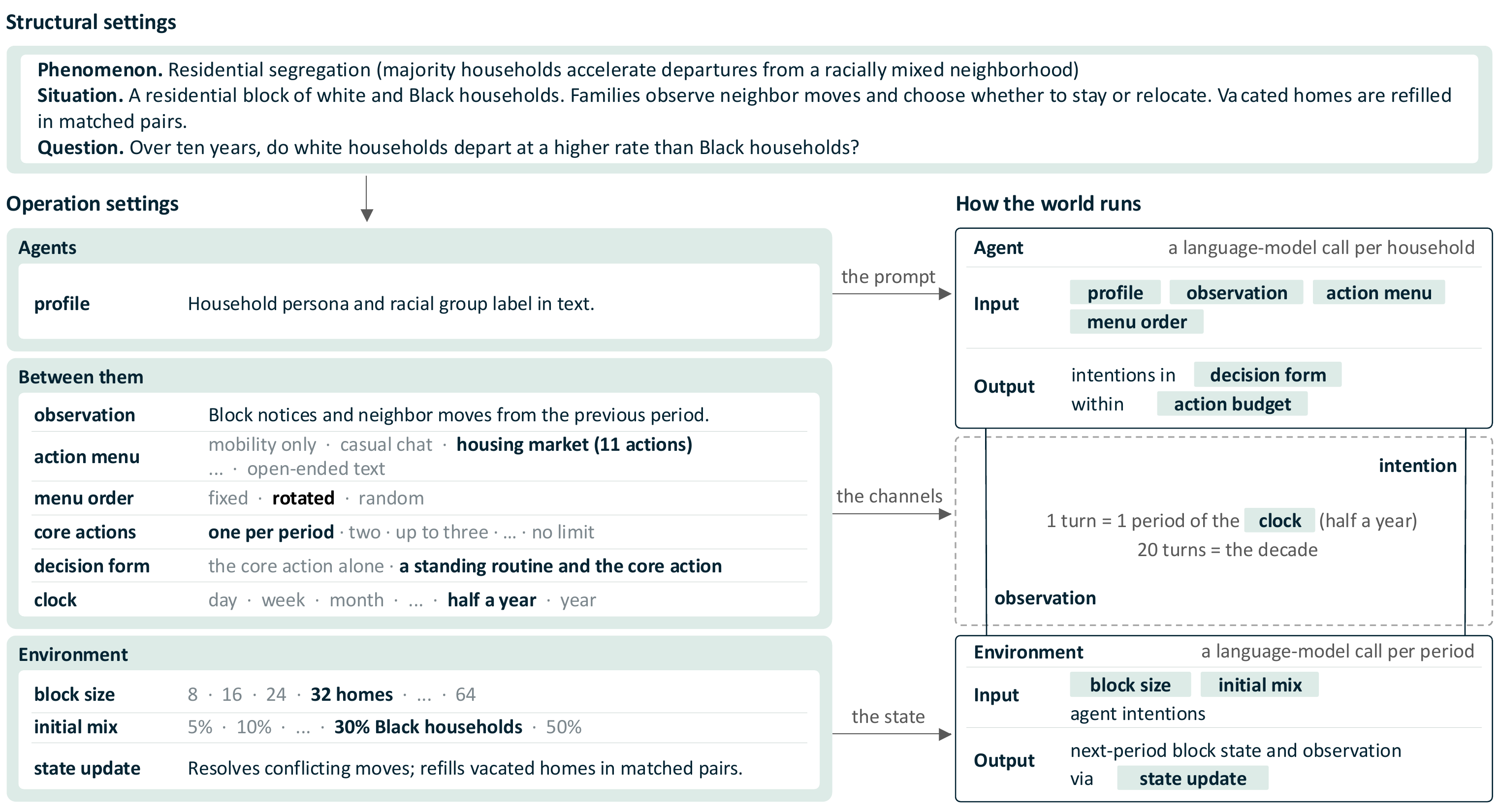}
    \caption{\textbf{Calibrating the baseline world}, illustrated with the residential segregation study in Section~\ref{sec:case}.
    The figure shows the world without a candidate rule. Top: Structural settings fix the social situation and research question. Left: Step~1 calibrates the \emph{operation settings} through baseline runs, varying one setting at a time. Grey values show alternatives, and bold values mark retained levels used in all further steps. Right: Agents express intentions, and the environment updates the world over twenty periods. In the further steps, the rule condition adds one candidate rule sentence to each agent's \emph{profile}. Both conditions use the same calibrated operation settings.}
    \Description{A baseline world for the residential segregation study, without a candidate rule. The top panel fixes the social situation and research question. The left panel lists operation settings calibrated in Step 1, with retained levels in bold. The right panel shows agent inputs and intentions, and environment updates over twenty periods. Candidate rules are not depicted. In the further steps, the rule sentence is appended to each agent's profile only in the rule condition.}
    \label{fig:world}
\end{figure}

We begin by constructing a simulation environment, which we call a \emph{world}. A world specifies the entities that participate in the simulation and the level at which they act (Figure~\ref{fig:world}). An agent need not represent an individual person. Depending on the research question, it may represent a household, an organization, a city, or even a nonhuman animal. This flexibility draws on the ability of language models to produce believable behavior from textual accounts of how such entities act, respond, and interact~\citep{park2023generative,zhou2024sotopia,shanahan2023roleplay}. The world also specifies what each agent can observe, what actions it can take, and how interactions unfold. Its temporal structure determines what counts as a simulation step. A step may represent an interval of clock or calendar time, such as a day or a year, or an interaction turn in which one or more agents observe and respond. These choices establish the units of action and time through which individual behavior gives rise to a collective outcome. These settings are of two kinds (Figure~\ref{fig:world}). \emph{Structural settings} characterize the social phenomenon under study and delimit the research question. \emph{Operation settings} translate this structure into a particular simulation design, such as the action menu, decision format, or block size. Step~1 calibrates the operation settings while holding the structural settings fixed.

Within this environment, we represent a hypothesized individual-level mechanism as a \emph{local rule}. The rule consists of a single behavioral instruction appended to each agent's profile. A local rule selected for evaluation is called a \emph{candidate rule}, and the mechanism is evaluated through the behavior produced by that rule. The \emph{baseline} uses the same world without the candidate rule. Comparing the candidate-rule condition with this baseline isolates the change in collective behavior associated with the hypothesized mechanism. Each run logs every prompt and response as its \emph{trace}, which records agents' actions and the resulting changes in the world.

We measure this change using a preregistered collective outcome, denoted by $Y$. The metric and the predicted direction of the candidate rule's effect are specified before the rule is tested~\citep{nosek2018prereg}. In our segregation study, $Y$ is the white households' departure rate minus the Black households' departure rate, where each rate is the group's departures over the run divided by the household-periods in which it was present, in percentage points per period. A positive value indicates that white households leave at a higher rate than Black households, and a larger value indicates a greater difference between the two groups (Section~\ref{sec:seg-method}).

\subsection{Step 1: calibrate and fix the baseline world}
\label{sec:phase1}

\begin{figure}[t]
\centering
\newcommand{\phaseitem}[2]{\par\vspace{2pt}\hangindent=1.5em\hangafter=1\noindent\textcircled{\scriptsize #1}\enspace #2}
\begin{tikzpicture}[
  font=\fontfamily{phv}\selectfont\footnotesize,
  phasebox/.style={draw=black, fill=white, line width=.9pt,
    inner xsep=6pt, inner ysep=6pt, align=left,
    text width=\dimexpr\textwidth/3-21pt\relax,
    anchor=north west, draw=none,
    execute at begin node={\hyphenpenalty=10000\exhyphenpenalty=10000}},
  flow/.style={draw=black, line width=1.2pt,
    -{Triangle[length=5.5pt,width=5pt]}},
  looplabel/.style={font=\fontfamily{phv}\selectfont\scriptsize\bfseries,
    fill=white, inner sep=2pt, align=center}
]

\node[phasebox] (plan) at (0,0) {%
  {\bfseries Register the plan}\par\smallskip
  \phaseitem{1}{Structural settings (fixed)}
  \phaseitem{2}{Operation settings: alternatives and calibration order}
  \phaseitem{3}{Baseline repetitions}
  \phaseitem{4}{Execution checks}
  \phaseitem{5}{Required actions and interactions}
  \phaseitem{6}{Room for the outcome to move in both directions}};

\node[phasebox] (verify) at ([xshift=10pt]plan.north east) {%
  {\bfseries Run and verify}\par\smallskip
  For each alternative:\par
  \phaseitem{1}{Run baseline repetitions}
  \phaseitem{2}{Apply execution checks}
  \phaseitem{3}{Review traces manually}\par\smallskip
  \vspace{4pt}Fault found: repair and rerun affected baselines.};

\node[phasebox] (choose) at ([xshift=10pt]verify.north east) {%
  {\bfseries Select an alternative}\par\smallskip
  Verified baselines must:\par
  \phaseitem{1}{Support required actions and interactions}
  \phaseitem{2}{Allow the outcome to increase and decrease}\par\smallskip
  If several qualify, retain the simpler alternative.};

\node[fit=(plan)(verify)(choose),inner sep=0pt] (row) {};
\foreach \n in {plan,verify,choose}
  \draw[line width=.9pt] (\n.north west) rectangle (\n.south east |- row.south);

\node[draw, line width=.9pt, inner sep=6pt, align=left,
  text width=\dimexpr\textwidth-15pt\relax, anchor=north west]
  (freeze) at ([yshift=-13mm]row.south west) {%
  {\bfseries Freeze settings and screen models}\par
  \phaseitem{1}{Freeze all operation settings after calibration.}
  \phaseitem{2}{Admit LLMs whose baselines pass the same checks, manual review, \mbox{and calibration criteria.}}
  \phaseitem{3}{Use this fixed world in the subsequent steps.}};

\draw[flow] ([yshift=-10pt]plan.north east) -- ([yshift=-10pt]verify.north west);
\draw[flow] ([yshift=-10pt]verify.north east) -- ([yshift=-10pt]choose.north west);
\coordinate (split) at ([yshift=-5mm]choose.south |- row.south);
\draw[line width=1.2pt] (choose.south |- row.south) -- (split);
\draw[flow] (split) -- (verify.south |- split) -- (verify.south |- row.south);
\node[looplabel] at ($(split)!0.5!(verify.south |- split)$)
  {next operation setting};
\draw[flow] (split) -- (split |- freeze.north);
\node[looplabel,anchor=east] at ([xshift=-3pt,yshift=-5mm]split)
  {all settings calibrated};

\end{tikzpicture}
\caption{\textbf{Step~1: calibrate and fix the baseline world.}
With structural settings fixed, researchers vary one operation setting at a time
and select an alternative using verified baseline runs. No candidate rule is
tested during calibration. A usable baseline supports the required actions and
interactions and leaves the outcome room to change in both directions; it need
not already exhibit the target collective pattern. Once all operation settings
are fixed, the same criteria screen the candidate LLMs for subsequent steps.}
\Description{Three top-aligned boxes show registration, baseline verification,
and alternative selection, each with circled items. Selection loops back to
verification for the next setting. Once all settings are calibrated, the flow
continues to a full-width box for freezing settings, screening models, and
using the fixed world in subsequent steps.}
\label{fig:phase1}
\end{figure}
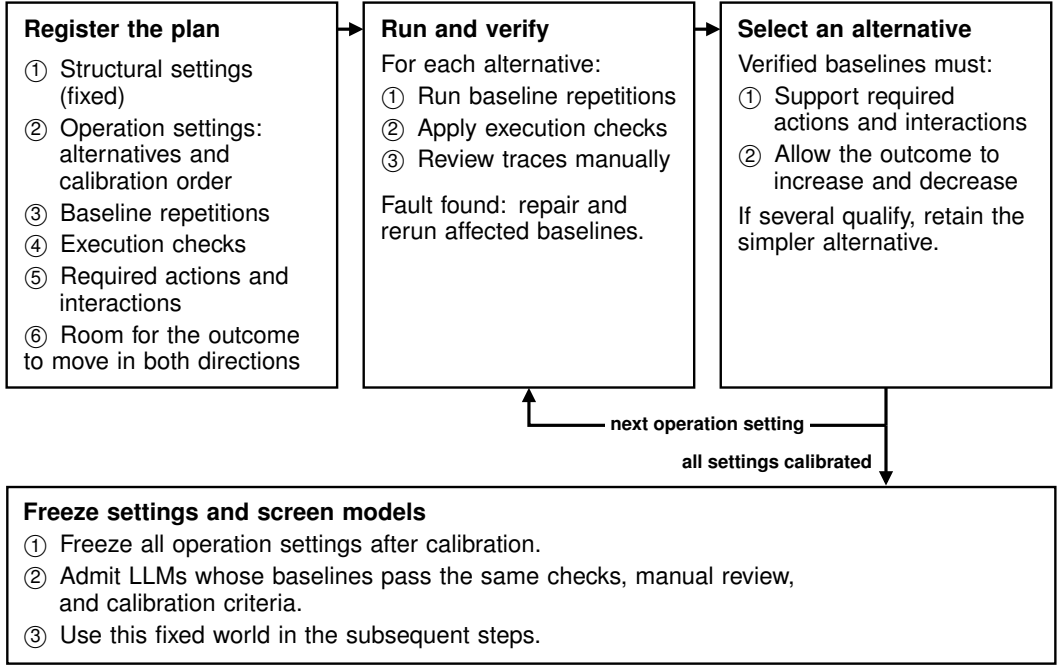

Step~1 establishes a functioning baseline world in which the actions and interactions needed for the target collective behavior can occur. It does not require the collective pattern itself to appear in the baseline; it ensures that the world can express it and that the outcome has room to change. To do so, it verifies each baseline while calibrating the world (Figure~\ref{fig:phase1}). Verification asks whether a run correctly implements the world being tested. Calibration uses only verified runs to select one alternative for each operation setting, while the structural settings remain fixed. Before any run, researchers specify the operation settings, their alternatives and calibration order, the number of baseline repetitions, researcher-written execution checks based on the world specification, the actions and interactions the baseline must support, and the required room for the outcome to move in both directions. They begin with one initial alternative for each operation setting.

%
%

The process begins with the initial baseline and then tests alternatives one operation setting at a time, holding all other settings fixed. For each alternative, it runs the registered baseline repetitions and automatically applies every registered execution check to each trace. Researchers also manually review the baseline traces. A failed check triggers a repair and rerun. If manual review reveals a fault not covered by the checks, researchers register a new check, repair the fault, and rerun every affected baseline. Existing checks are never removed or relaxed. Only runs that pass both the checks and manual review proceed to calibration.

An alternative qualifies when the relevant actions and interactions occur in its verified baselines and the outcome has room to move in both directions. This requirement serves the research question rather than a preferred result. No candidate rule is run during calibration, so calibration cannot select on any rule's effect. For example, in the segregation study, some households must be willing and able to move; if nobody moves, differential flight cannot emerge. The baseline need not already exhibit a large departure-rate gap, but the gap must be able to increase or decrease under the candidate rules. If several alternatives qualify, the method retains the simpler implementation. It then moves to the next operation setting and repeats the process. Once all operation settings are calibrated, they are frozen and the same requirements are used to screen the model pool.

Further steps estimate rule effects in this fixed world; discovering a new execution fault would reopen Step~1 rather than permit an unrecorded change during rule testing. Each study reports its calibration repetitions and acceptance criteria. When verification reveals a fault, it also reports the added check, repair, and affected reruns.

\subsection{Step 2: operationalize candidate mechanisms}
\label{sec:phase2}

Step~2 turns each candidate mechanism into a rule that can be added to the calibrated world. Rules come from the source literature or are formulated by the researchers.
Each rule is registered with a hypothesis about its effect on $Y$: an amplifying rule predicts
more of the pattern and a suppressive rule predicts less. We write $\sigma_r=+1$ for the first and
$\sigma_r=-1$ for the second.

In an LLM-agent simulation, an agent receives a rule as a prompt sentence and interprets it, and the same mechanism can be expressed in many different sentences. Language models are sensitive to surface phrasing~\citep{sclar2024quantifying,mizrahi2024state}, and a single sentence therefore measures the effect of that particular wording under the model's own priors. Running that one sentence many times does not separate the mechanism from the wording that carried it~\citep{clark1973language}. The method therefore evaluates each rule across many prompt wordings and across the admitted models. A \emph{configuration} for rule $r$ fixes the model and the three prompt texts that its run requires:
\begin{equation}
\begin{aligned}
  \texttt{configuration}_{ir} = \bigl(&\texttt{Model}_i,\; \texttt{Rule\ wording}_{ir},\\
  &\texttt{Profile\ wording}_i,\; \texttt{World\ wording}_i\bigr).
\end{aligned}
  \label{eq:config}
\end{equation}

\begin{figure}[!htbp]
 \includegraphics[width=\textwidth]{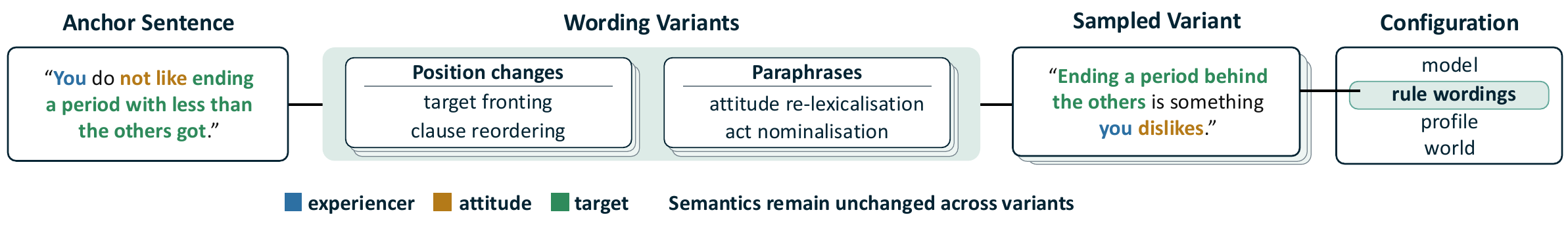}
  \caption{\textbf{Testing several expressions of the same rule.}
  We begin with an anchor sentence and rewrite it while preserving who holds
  the attitude, what the attitude is, and what it concerns. After these checks,
  the rewrites form a wording pool from which each configuration selects one
  sentence to append to every agent's profile. We then compare the resulting
  run with a baseline that uses the same model and background prompts but omits
  the rule sentence. Repeating this comparison assesses how the estimated effect varies
  across different expressions of the same rule.}
  \Description{A diagram from left to right: an anchor sentence with its
  experiencer, attitude and target color-coded; a variant produced by one
  grammatical operation; and a stack of accepted wordings forming the pool.}
  \label{fig:rule-wording}
\end{figure}

Each rule starts from an anchor sentence, taken from the literature or written by the researchers (Figure~\ref{fig:rule-wording}). The anchor fixes three aspects of the proposed mechanism: its \emph{experiencer}, who holds the attitude; its \emph{attitude}, what kind of evaluation it is; and its \emph{target}, what the attitude is about~\citep{eagly1993psychology,martin2005language}. Variants are prepared before their evaluation and checked to preserve all three. Two sentences with the same meaning can still draw different responses from a language model~\citep{elazar2021measuring}, so averaging across a declared pool supports a claim beyond one sentence. To assess whether that pool is large enough, we report wording uncertainty, $\text{SD}(\delta_{\text{wording}})/\sqrt{n_{\text{wording}}}$, where $\delta_{\text{wording}}$ is the estimated effect for one wording and $n_{\text{wording}}$ is the number of wordings. Section~\ref{sec:pg-wording} evaluates this quantity using subsets of a fixed pool of twenty-six wordings. This analysis measures the sensitivity of the pooled estimate; it does not select wordings for producing a preferred effect.

We also use a language model to paraphrase the profile and world prompts, keeping participant details, available actions, numerical settings, and interaction procedures unchanged.
\subsection{Step 3: estimate and compare rule effects}
\label{sec:rule-ranking}

Changing the model or background wording can change agent behavior even when the simulation parameters stay fixed. We therefore compare each rule with a baseline using the same model, profile text, world description, initial conditions, and any scheduled interactions. The rule condition appends the candidate-rule sentence to every agent's profile; the baseline omits it. We calculate the difference within each configuration before averaging across configurations:
\begin{equation}
  \delta_i(r) \;=\; \sigma_r\bigl[\,Y_i(r)-Y_i(0)\,\bigr],
  \qquad
  \widehat{\Delta Y}_r \;=\; \frac{1}{N}\sum_{i=1}^{N} \delta_i(r).
  \label{eq:ate}
\end{equation}
Signing by $\sigma_r$ ensures that a positive \emph{paired difference} $\delta_i(r)$ indicates an outcome moving in the predicted direction, and its average $\widehat{\Delta Y}_r$ is the \emph{rule effect}, reported in the metric's own units. Pairing does not eliminate stochastic variation or interactions between a rule and its context. Rule wording and sentence position have no counterpart in the baseline, so Section~\ref{sec:pg-wording} examines them separately.

Each study reports the configurations evaluated, the repetitions per condition, and how failed or incomplete runs were handled.
Each candidate rule is characterized by two quantities. The rule effect $\widehat{\Delta Y}_r$ measures its average signed change in the outcome. The standard deviation of its paired differences, $\mathrm{SD}(\delta_i(r))$, describes how much that change varies from one configuration to the next.

We compute 90\% intervals by bootstrapping whole configurations, keeping each rule run paired with its baseline. Because effects are signed by their registered predictions, an interval entirely above zero supports an effect in the predicted direction. To compare magnitude and consistency, we compute \emph{control power}:
\begin{equation}
  \mathrm{CP}_r \;=\;
  \frac{\widehat{\Delta Y}_r}{\mathrm{SD}\!\left(\delta_i(r)\right)}
  \label{eq:cp}
\end{equation}
Control power coincides with the paired standardized effect $d_z$~\citep{lakens2013calculating}. It is an effect-size measure, not statistical power. We rank rules by their mean signed effect relative to its variation across configurations. This ranking compares intervention effects; it does not rank the validity of the proposed explanations.

A ranking also has sampling uncertainty. We assess its stability by repeating the analysis on resampled configurations and comparing effect estimates and rule order with the full-sample results. The full sample is a reference for this analysis, not a known ground-truth ranking. Section~\ref{sec:pg-cost} reports this analysis (Figure~\ref{fig:pg-convergence}). More configurations can improve precision, but they do not guarantee that two similar rules become distinguishable.

\subsection{Step 4: assess explanatory support from the behavioral pathway}
\label{sec:process-evidence}

A micro-to-macro explanation proposes how individual behavior and interaction produce a collective outcome. Adding a candidate rule tests an implication of that account: whether the outcome changes in the predicted direction. The matched comparison estimates the effect of adding the rule sentence, not the causal effect of each intermediate behavior. Behavioral records help assess the proposed link by showing whether the relevant behavior occurs and whether the sequence of interactions and outcomes is consistent with the account.

Our studies examine different links according to the phenomenon: group-specific departures in segregation, punishment and subsequent contribution in public goods, discussion and opinion change in polarization, and contributions before exposure to social information in cascades. These analyses distinguish a missing outcome effect from a missing or unsupported behavioral link. Their interpretation also depends on what the metric measures: a rule can change contribution amounts without increasing contribution frequency.

Trace evidence can reveal a gap in an explanation, but a compatible trace does not establish that the proposed process is necessary or uniquely responsible for the effect. For example, an effect present before agents observe others cannot be attributed to that observation, although later social transmission remains possible. Testing that later process requires an additional intervention on the information channel. We therefore distinguish the registered outcome comparison from behavioral analyses that qualify its interpretation and motivate follow-up experiments, rather than treating these analyses as a separate causal identification of the mechanism.

\section{Evaluation Design}
\label{sec:evaluation-overview}

We create four simulation worlds to evaluate different parts of the method. The worlds are grounded in established social-science models or experiments and cover residential sorting, repeated contribution decisions, group discussion, and cooperation across changing partners. They provide complementary tests rather than four equivalent replications.

\noindent\textbf{RQ1: Operationalization.}
The residential-segregation study demonstrates the complete workflow. We examine whether calibration produces an executable world with a responsive outcome, and whether matched interventions distinguish the direction, magnitude, and consistency of eight candidate-rule effects. (Section \ref{sec:case})

\noindent\textbf{RQ2: Reliability.}
The public-goods study examines when a rule comparison can be relied on. Using 268 configurations across four models, we assess how effect conclusions and rankings change as configurations accumulate, and how estimates vary with rule wording and sentence position. Agreement with the full-sample analysis measures stability, not accuracy against a known true ranking. (Section \ref{sec:pg})

\noindent\textbf{RQ3: Explanatory support.}
Across all four studies, we examine what behavioral records add to an outcome comparison. The segregation and public-goods studies distinguish changes in the registered outcome from changes in underlying behaviors. The group-polarization and cooperative-cascade studies provide focused tests of whether the behavior named by a rule is followed by the predicted collective change and whether that change begins before or after the proposed social interaction. (Section \ref{sec:four})



\FloatBarrier
\section{Operationalization}
\label{sec:case}

To address RQ1, we use residential segregation, a classic example of collective emergence~\citep{schelling1971dynamic}, to demonstrate how the method establishes and quantifies comparisons among candidate rules. A family's decision to move changes the neighborhood seen by other families. As these decisions accumulate, a mixed neighborhood can become segregated. In U.S. census data, Card et al.~\citep{card2008tipping} found that white population growth dropped sharply in neighborhoods whose minority share lay just above a city-specific threshold. This section reports whether calibration yields an executable world with a responsive outcome, and whether eight mechanisms, once operationalized as rules, produce effects that can be measured and compared.

A neighborhood's changing population does not, by itself, reveal households' reasons for staying or leaving. A household may prefer neighbors of its own group~\citep{clark1991residential}, expect similar neighbors to leave~\citep{ellen2000sharing}, or feel attached to its home and street~\citep{brown2003place}. We represent these and other candidate mechanisms as eight candidate rules (Table~\ref{tab:rule-slate}). We treat them as alternative micro-level explanations for differential residential sorting and test which rules increase or reduce the difference between white and Black households' departure rates, and how consistently they do so.

\subsection{World and outcome}
\label{sec:seg-method}

\textbf{The simulated block.}
The world represents a residential block in mid-century America. Each agent represents a household and receives a text profile that includes its racial group label. Households receive information about their neighbors and recent departures, then express intentions such as interacting with neighbors, considering housing options, moving, or carrying on as usual. The environment resolves these intentions and updates the block. Vacated homes are refilled in pairs, one white household and one Black household, so that relocation can continue without emptying the block. Because arrivals are balanced, differences in departure rates drive changes in the block's racial composition. Figure~\ref{fig:world} shows the operation settings used to implement this situation; calibration selects their values.\\
\textbf{Outcome and baseline.}
The primary outcome is the \emph{departure-rate gap}: the percentage of white households leaving minus the percentage of Black households leaving, pooled over the decade and reported in percentage points per period. For example, rates of 6\% and 3\% give a gap of 3 percentage points. Zero means equal departure rates; a positive value means white households leave at a higher rate. Within each configuration, we divide each group's total departures across the twenty periods by the number of household-periods in which that group was present. We then subtract the Black departure rate from the white departure rate. The baseline is the same simulated world without a candidate rule. An amplifying rule is predicted to widen its departure-rate gap, and a suppressive rule is predicted to narrow it.

\subsection{Calibrating the baseline world}
\label{sec:establish}

\textbf{Method.}
We ran baseline simulations on DeepSeek V4 Flash to compare operation settings, varying one setting at a time while holding the others fixed. Figure~\ref{fig:world} shows the range of values for each operation setting and highlights those retained. We calibrated the action budget first. The menu order came next, followed by the decision form and then the block size. Each tested value received six baselines. For each value, we applied the current execution checks and reviewed all six traces by hand. In our runs, trace review revealed that some decisions to move were not executed by the environment. We recorded the fault, added a move-execution check, repaired the environment, and repeated the affected comparisons. In all, calibration went through three rounds of verification before the selected settings passed all checks.

Three execution checks were in force by the end of calibration. Every recorded action had to come from the action menu. Every household had to keep its original racial-group label. Every decision to move had to be executed. A baseline that failed a check was repaired and rerun. The outcome requirement was room for the departure-rate gap to move in both directions. We set the threshold at one-fifth of the block, about seven departures per baseline on average over the six verified baselines of an alternative. The departure threshold ensures that relocation occurs in the baseline world. Once the last operation setting was chosen, we froze the operation settings and screened the model pool in the fixed world with the same execution checks and departure threshold. Four models were admitted (Appendix~\ref{app:models}).

\textbf{Results.}
The action-budget comparison illustrates how calibration affected the baseline. Under \emph{pick as many or as few as the situation calls for}, households spent their turns on conversation and routine activities and rarely reached a decision to move. Under \emph{pick one action}, relocation decisions appeared in all six baselines, compared with one of the six baselines under the open-ended budget. The one-action setting was retained for rule testing.

The resulting world contains 32 homes, starts with a Black population share set to 30\%, and runs for twenty half-year periods. Each household describes its routine and chooses one core action from an eleven-action menu in each period. These settings, together with the observation and refill procedures, remain fixed in the further steps.

\subsection{Registering and testing candidate rules}
\label{sec:rules}

With the operation settings fixed, we test how adding each candidate rule changes relocation relative to the baseline. We compare both the size of that change and its consistency across configurations.

\textbf{Method.}
We test eight candidate rules, listed with their sources and sentences in Table~\ref{tab:rule-slate}. Each sentence is appended to every household's profile. Four rules predict a wider departure-rate gap: minority-status aversion, own-group attraction, out-group prejudice, and expectation of flight. Four predict a narrower gap: place attachment, intergroup contact, integration preference, and collective efficacy.

Minority-status aversion, for example, describes a household's preference about the group composition of its neighbors. The model interprets this sentence in the household's context; the simulation does not impose a numerical threshold at which it must move. The initial 30\% Black share describes the block's composition and does not make white households a numerical minority.

We evaluate twelve configurations spanning the four admitted models and alternative phrasings of the rules, household profiles, and world description, using the paired comparisons defined in Section~\ref{sec:rule-ranking}. The configuration allocation is three per admitted model, drawn from five registered rule wordings, five household-profile wordings, and four world-description wordings; Table~\ref{tab:rule-slate} shows the anchor sentence for each rule-wording pool. We run one baseline and one run for each of the eight rules in every configuration, for 108 completed runs. Calibration runs were not reused; each rule run was paired with a newly collected baseline under the same configuration, and no outcome-based exclusions were applied.

For each rule, we calculate the signed paired differences in the gap and their mean (Equation~\eqref{eq:ate}). We report effects in percentage points and rank rules within the amplifying and suppressive groups by control power. The 90\% intervals come from bootstrapping whole configurations while preserving each rule--baseline pair. We draw 10,000 resamples, and each interval is the estimate plus or minus 1.65 bootstrap standard errors. Group-specific departure rates, total departures, and departure timing describe how behavior changed in the same runs.

\begin{figure}[!htbp]
  \centering
  \includegraphics[width=\linewidth]{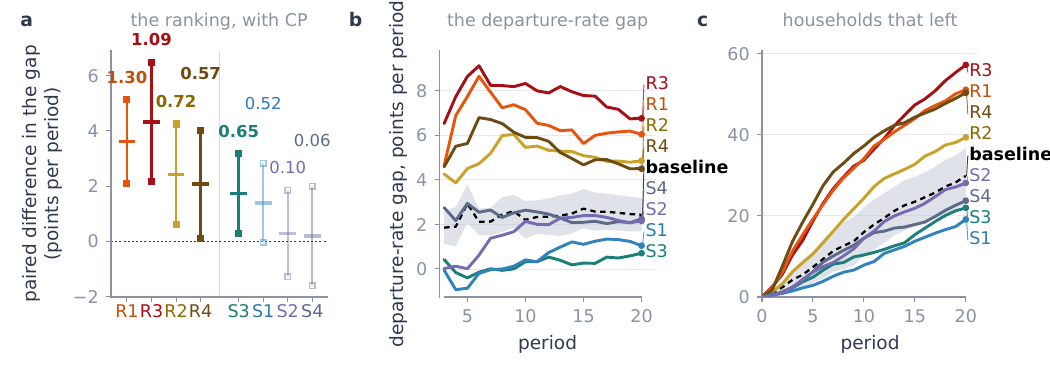}
  \caption{\textbf{How each rule changes relocation in the same simulated block.}
  \textbf{(a)} Horizontal marks show each rule's mean paired difference across
  the twelve configurations and vertical bars its 90\% interval; solid
  intervals lie entirely in the registered direction. Effects
  are signed by the predicted direction, so a positive value means a wider gap for an amplifying rule and
  a narrower gap for a suppressive rule. Numbers above the intervals are
  control power (CP), which combines effect size and consistency.
  \textbf{(b)} The white-minus-Black departure-rate gap, computed from all
  departures up to each period and averaged across configurations.
  \textbf{(c)} Cumulative departures, averaged across configurations.
  Dashed lines show the baseline and shaded bands show one standard error
  around its mean. The trajectories describe how the block changes; the paired
  differences in (a) measure the added rule's effect.}
  \Description{The figure has three panels, described from left to right. The
  left panel shows the paired difference in the gap for each of the eight
  rules, identified by their rule codes: a bar at the
  mean and a whisker spanning the 90 percent interval, the amplifying rules to
  the left of a vertical line and the suppressive rules to its right, and the
  control power printed above each rule. Rules whose intervals lie entirely in
  the registered direction are drawn solid and the rest faded. The middle panel plots the departure-rate
  gap from period three to period twenty under all eight rules, each in its own
  color and marked at the end of its line with its rule code, around a dashed
  baseline line inside a grey band of one standard error; the four amplifying
  rules run well above the baseline for the whole decade and the four
  suppressive rules run at or below it. The right panel plots the cumulative
  number of households that have left the block under each rule; R3 ends
  highest near fifty-seven, R1 and R4 end together near fifty, R4 rising
  fastest in the first six periods, the baseline is a nearly straight line
  ending near thirty, and S1 ends lowest near nineteen.}
  \label{fig:seg-result}
\end{figure}

\begin{table}[t]
  \centering
  \caption{The eight registered candidate mechanisms. Each rule
  enters the simulation as one sentence appended to every household's profile,
  and nothing else in the world changes between conditions.}
  \label{tab:rule-slate}
  \small
  \renewcommand{\arraystretch}{1.14}
  \begin{tabular}{@{}
      >{\raggedright\arraybackslash}p{0.04\columnwidth}
      >{\raggedright\arraybackslash}p{0.30\columnwidth}
      >{\raggedright\arraybackslash}p{0.60\columnwidth}@{}}
    \toprule
    & Mechanism & Sample injected sentence \\
    \midrule
    \multicolumn{3}{@{}l}{\emph{Amplifying: the mechanism's own literature predicts more sorting}} \\[2pt]
    R1 & Minority-status aversion \rulesource{schelling1971dynamic,farley1978chocolate}
       & ``Your family would rather not live where most of the neighbors are
         not your own kind.'' \\[2pt]
    R2 & Own-group attraction \rulesource{clark1991residential,bruch2006neighborhood}
       & ``Your family would rather live where most of the neighbors are your own kind.'' \\[2pt]
    R3 & Out-group prejudice \rulesource{bobo1996attitudes}
       & ``Your family would rather not live beside families unlike
         yourselves.'' \\[2pt]
    R4 & Expectation of flight \rulesource{ellen2000sharing,granovetter1978threshold}
       & ``Your family expects that neighbors like you will leave before
         long.'' \\[4pt]
    \multicolumn{3}{@{}l}{\emph{Suppressive: a separate literature predicts less}} \\[2pt]
    S1 & Place attachment \rulesource{fried2000continuities,brown2003place}
       & ``Your family feels that this house and this street are part of who you are.'' \\[2pt]
    S2 & Intergroup contact \rulesource{allport1954nature,pettigrew2006meta}
       & ``Your family finds that getting to know new neighbors leaves you
         thinking better of them.'' \\[2pt]
    S3 & Integration preference \rulesource{krysan2002residential}
       & ``Your family wants its children to grow up among different kinds of
         people.'' \\[2pt]
    S4 & Collective efficacy \rulesource{sampson1997neighborhoods}
       & ``Your family counts on the neighbors to look out for the block, and
         they count on you.'' \\
    \bottomrule
  \end{tabular}
  \Description{The table has eight rows listing the candidate rules in two
  groups. The four amplifying rules are minority-status aversion, own-group attraction,
  out-group prejudice, and expectation of flight. The four suppressive rules are
  place attachment, intergroup contact, integration preference, and collective
  efficacy. Each row gives the mechanism name, its literature sources, and the
  exact sentence injected into each household's profile.}
\end{table}
\textbf{Results.}
\label{sec:seg-ranking}
Across the twelve configurations, the baseline white departure rate is 5.8\% per period and the Black rate is 3.4\%, giving a gap of 2.4 percentage points (Table~\ref{tab:phase2-ranking}). Figure~\ref{fig:seg-result} shows both the paired effects and how relocation develops over the decade.

All four amplifying rules widen the gap, providing intervention-effect evidence for these candidate mechanisms under the tested simulation conditions. The gap is not a sum of isolated decisions: each period a household observes its neighbors' departures and can talk with a neighbor, ask around, or pass on what it heard, so what one family learns or plans can reach others before anyone moves.
Minority-status aversion (R1) and out-group prejudice (R3) increase departures
in both groups, with a larger increase among white households. R3 produces
the most departures and the largest mean effect on the gap. It ranks below
R1 because its effect varies more across configurations. Own-group attraction
(R2) uses the same sentence for both groups, but primarily increases white
departures; the Black rate remains close to baseline. Expectation of flight
(R4) has the smallest mean effect among the amplifying rules because Black
departures also increase substantially. Its traces show earlier departures:
the mean departure occurs at period 7.1, compared with 10.6 in the baseline.
This timing difference is consistent with accounts in which expectations of
others' departures influence when households move~\citep{ellen2000sharing,granovetter1978threshold}.

\begin{table}[t]
  \centering
  \caption{Rule effects on the departure-rate gap, ranked within each family by
  control power (Equation~\eqref{eq:cp}). The departure rate is a group's departures
  over the decade divided by its household-periods, in percent, and the gap is the white rate minus
  the Black rate. $\widehat{\Delta Y}_r$ is the rule effect, the mean paired
  difference in the gap in points per period, signed by the registered
  direction. $\downarrow$ marks rules registered to narrow the gap. Effects are computed from unrounded rates.}
  \label{tab:phase2-ranking}
  \small
  \setlength{\tabcolsep}{4.5pt}
  \begin{tabular}{@{}llrrrrr@{}}
    \toprule
    & & \multicolumn{3}{c}{Departure rate, \% per period} & \multicolumn{2}{c}{Effect on the gap} \\
    \cmidrule(lr){3-5} \cmidrule(lr){6-7}
    & Rule & White & Black & Gap & $\widehat{\Delta Y}_r$ & $\mathrm{CP}_r$ \\
    \midrule
    & baseline (no rule sentence) & 5.8 & 3.4 & $+2.4$ & & \\
    \multicolumn{7}{@{}l}{\textbf{Amplifying}} \\
    1 & R1 minority-status aversion & 11.2 & 5.2 & $+6.0$ & $+3.6$ & $\mathbf{+1.30}$ \\
    2 & R3 out-group prejudice      & 12.7 & 5.9 & $+6.8$ & $+4.3$ & $+1.09$ \\
    3 & R2 own-group attraction     &  8.7 & 3.8 & $+4.9$ & $+2.4$ & $+0.72$ \\
    4 & R4 expectation of flight    & 10.4 & 5.9 & $+4.5$ & $+2.1$ & $+0.57$ \\
    \multicolumn{7}{@{}l}{\textbf{Suppressive}} \\
    1 & S3 integration preference $\downarrow$ & 3.8 & 3.1 & $+0.7$ & $+1.7$ & $\mathbf{+0.65}$ \\
    2 & S1 place attachment $\downarrow$       & 3.5 & 2.4 & $+1.0$ & $+1.4$ & $+0.52$ \\
    3 & S2 intergroup contact $\downarrow$     & 5.4 & 3.3 & $+2.2$ & $+0.3$ & $+0.10$ \\
    4 & S4 collective efficacy $\downarrow$    & 4.8 & 2.5 & $+2.2$ & $+0.2$ & $+0.06$ \\
    \bottomrule
  \end{tabular}
  \Description{The table has a baseline row and two blocks of four rules,
  amplifying and suppressive, each ranked by control power. For every row it
  gives the white and the Black departure rate and their gap, then the rule
  effect in points per period and the control power. The
  baseline rates are 5.8 and 3.4 percent a period, a gap of 2.4.
  Minority-status aversion leads the amplifying block with control power 1.30
  and a gap of 6.0; out-group prejudice has the widest gap, 6.8, and the
  largest effect, 4.3 points, but ranks second at 1.09; own-group attraction
  and expectation of flight follow.
  Integration preference leads the suppressive block at 0.65 with the two rates
  closest together, 3.8 against 3.1; place attachment follows at 0.52, and
  intergroup contact and collective efficacy leave the gap where it was.}
\end{table}

Of the suppressive rules, integration preference (S3) has an established effect.
It lowers the white departure rate to within one percentage point of the Black
rate. Place attachment (S1) has a smaller positive estimate, with its interval
reaching zero. It lowers departures in both groups and delays the mean departure
to period 13.5, compared with 10.6 in the baseline. Households spend 58\% of
their periods on routine activities, the largest share under any candidate rule.

Intergroup contact (S2) and collective efficacy (S4) do not establish effects
on the gap. Contact theory proposes that interaction can improve attitudes
toward other groups~\citep{pettigrew2006meta}. Under S2, however, households direct
34\% of their core actions toward neighbors, compared with 43\% in the baseline.
The traces therefore do not show the increased contact expected from this
account. S4 lowers both group departure rates by similar amounts, leaving
little change in their difference. These cases illustrate why the primary
metric and the behavioral traces need to be interpreted together.\\

\section{Reliability}
\label{sec:pg}

To address RQ2, our public-goods study, adapted from Fehr and G\"{a}chter~\citep{fehr2000cooperation}, evaluates the reliability of the method's quantitative comparisons, focusing on how effect estimates and rankings stabilize as more configurations are included and how effects vary with rule wording and sentence position. Behavioral records further show whether changes at the punishment and contribution stages align with the proposed process. The full-sample results on all 268 configurations provide the reference for these analyses.

\subsection{World setup and calibration}
\label{sec:pg-world}

The public-goods world has four participants at one table. The game allows ten periods. In each period, every participant receives twenty credits (units of game currency) and decides how many to put into a shared project and how many to keep. Every contributed credit returns 0.4 credits to each member of the group, which generates a collective return of 1.6 credits and imposes a personal net loss of 0.6 credits on the contributor. Free-riding is therefore individually optimal, while full contribution maximizes collective welfare. Once the contributions are shown, a second stage allows each participant to punish another by assigning deduction points, at a price the participant pays. The first point costs the sender one credit and takes two from the target, and further points follow the price list of the original study. In no period does a participant learn who punished them.

The collective pattern is whether the group keeps contributing over the
periods or slides toward everyone keeping their credits. In the human
experiment, contributions decay when nobody can punish and stay high when
participants can punish one another~\citep{fehr2000cooperation}. Our world keeps
the punishment stage because several registered mechanisms work through it.
The registered outcome $Y$ is the percentage of participant-periods in which
the contribution standing at the period's close is greater than zero. A higher value means that they contribute in a larger share of
their decisions. We also report the average number of credits put
in and the number of punishment acts, which show how a rule changed behavior
whatever its effect on $Y$.

\textbf{Baseline calibration.}
\label{sec:pg-calibration}
Calibration selects the operation settings and screens the model pool using baseline runs. One calibration comparison varied how the participant's task was stated. An instruction to maximize final earnings left contributions near zero, leaving little room to test suppressive rules. A neutral instruction to decide how to use the available credits was retained instead. Four models are admitted (Appendix~\ref{app:models}), with baseline outcomes that leave room for changes in both directions. The selected operation settings remain fixed throughout the rule comparisons and reliability analyses below.

\subsection{Rule effects and full-sample ranking}
\label{sec:pg-effects}

We then test which candidate rules change contribution frequency and how consistently they do so across configurations. We report the full-sample effects and rankings here as the reference for the reliability analyses that follow.

\textbf{Method.} We test five mechanisms from behavioral economics (Table~\ref{tab:pg-slate}). Inequity aversion (R1), conditional cooperation (R2), and negative reciprocity (R3) predict more cooperation. Money maximization (R4) and removing the shadow of the future (R5) predict less. Each rule is added as one sentence at the end of every participant's profile and has twenty-six registered wordings.

We obtained complete baseline and rule records for 268 configurations. Each of the four admitted models contributes 67 configurations. Every configuration contains one baseline and one run for each of the five rules, for 1,608 runs in total. We estimate each rule's signed paired effect on contribution frequency and rank rules by control power, following Section~\ref{sec:rule-ranking}. Because runs can complete different numbers of periods, we also repeat the comparison on pairs matched by completed periods.

\begin{table}[!htb]
  \centering
  \caption{The registered rules for the public-goods world. Each rule enters as
  one sentence appended to every participant's profile; the sentence shown is
  the registered anchor, and each rule has twenty-six wordings of it. Rules
  marked $\downarrow$ are registered as suppressive.}
  \label{tab:pg-slate}
  \small
  \renewcommand{\arraystretch}{1.14}
  \begin{tabular}{@{}
      >{\raggedright\arraybackslash}p{0.04\columnwidth}
      >{\raggedright\arraybackslash}p{0.31\columnwidth}
      >{\raggedright\arraybackslash}p{0.59\columnwidth}@{}}
    \toprule
    & Mechanism & Sample injected sentence \\
    \midrule
    R1 & Inequity aversion \rulesource{fehr1999theory}
       & ``You do not like ending a period with less than the others got.'' \\[2pt]
    R2 & Conditional cooperation \rulesource{fischbacher2001people,fischbacher2010social}
       & ``You put in about what you expect the others to put in.'' \\[2pt]
    R3 & Negative reciprocity \rulesource{fehr2000cooperation,fehr2002altruistic}
       & ``You are willing to spend to make a point, even when it costs
         you.'' \\[2pt]
    R4 & Money-maximizing $\downarrow$ \rulesource{ledyard1995public}
       & ``You take the option that leaves you the most credits.'' \\[2pt]
    R5 & Shadow of the future, removed $\downarrow$ \rulesource{andreoni1988why}
       & ``You will not meet these people again.'' \\
    \bottomrule
  \end{tabular}
  \Description{A five-row table of the public-goods rules. The rules are
  inequity aversion, conditional cooperation, negative reciprocity, a
  money-maximizing rule and a rule removing the shadow of the future. Each row
  gives the mechanism, its literature source, and the registered sentence
  appended to every participant's profile. The last two rules are registered to
  lower contribution.}
\end{table}

\textbf{Results.}
In the baseline, participants contribute in 84.8\% of their decisions. Contribution frequency falls from 88.6\% over the first three periods to 79.2\% over the last three, despite the available punishment stage. Participants who had just been punished contributed no more in the following period than those who had not, and 52\% of punishment targeted participants already contributing at or above their table's average. These observations do not show the pattern of sustained cooperation reported in the human punishment condition.

Against this baseline, money maximization (R4) is the strongest suppressor ($\mathrm{CP}=2.90$), and conditional cooperation (R2) is the strongest amplifier ($\mathrm{CP}=0.59$). Removing the shadow of the future (R5) has a weaker suppressive effect ($\mathrm{CP}=0.38$). Inequity aversion (R1) does not establish an effect, while negative reciprocity (R3) has an established effect opposite to its registered prediction. Table~\ref{tab:pg-summary} reports the full-sample effects and rankings.

\begin{table}[!htb]
  \centering
  \caption{The public-goods rules, $N=268$ configurations, ranked by control power (Equation~\eqref{eq:cp}). Every participant makes one
  contribution decision per period. In the baseline, 84.8\% of them choose to
  contribute. $\widehat{\Delta Y}_r$ is the change from baseline in percentage
  points, signed by the predicted direction; positive values mean less
  contribution for suppressive rules. SE is the standard error of this effect.
  $\downarrow$ marks the rules registered as suppressive.
  Punish is the change in the number of punishment acts a run, against 6.7 in
  the baseline. Credits is the change in the average number of credits
  contributed.}
  \label{tab:pg-summary}
  \small
  \setlength{\tabcolsep}{4pt}
  \begin{tabular}{@{}llrrrrr@{}}
    \toprule
    & Rule & $\widehat{\Delta Y}_r$ & SE & $\mathrm{CP}_r$ & Punish & Credits \\
    \midrule
    \multicolumn{7}{@{}l}{\emph{Amplifying}} \\
    1 & R2 conditional cooperation & $+12.0$ & 1.3 & $\mathbf{+0.59}$ & $-3.6$ & $+0.98$ \\
    2 & R1 inequity aversion & $+0.4$ & 1.7 & $+0.01$ & $+4.6$ & $+1.26$ \\
    3 & R3 negative reciprocity & $-2.7$ & 1.5 & $-0.11$ & $+15.6$ & $+1.45$ \\
    \midrule
    \multicolumn{7}{@{}l}{\emph{Suppressive}} \\
    1 & R4 money-maximizing $\downarrow$ & $+78.6$ & 1.7 & $\mathbf{+2.90}$ & $-4.9$ & $-7.61$ \\
    2 & R5 shadow removed $\downarrow$ & $+11.5$ & 1.8 & $+0.38$ & $+0.8$ & $-0.93$ \\
    \bottomrule
  \end{tabular}
  \Description{A five-row table in two blocks, ranking the public goods rules by
  control power inside each family. The amplifying block holds conditional
  cooperation, inequity aversion and negative reciprocity; the suppressive block
  holds the money-maximizing rule and the removed shadow of the future. Each row
  gives the rule effect in percentage points, its standard error, its control
  power, and two further readings: the change in punishment acts a run and the
  change in average contribution.}
\end{table}

Money maximization (R4) lowers contribution frequency by 78.6 percentage points
and nearly eliminates punishment. Conditional cooperation (R2) raises the
frequency by 12.0 points even as punishment falls below baseline. Removing the
shadow of the future (R5) lowers it by 11.5 points while punishment remains near
baseline.

R1 and R3 substantially change punishment behavior. Inequity aversion (R1) adds 4.6 punishment acts per run
without reliably changing contribution frequency. Negative reciprocity (R3),
the mechanism proposed by Fehr and G\"{a}chter~\citep{fehr2000cooperation}, adds 15.6 punishment
acts, yet contribution frequency falls by 2.7 points. Increased punishment is consistent with the behavior named by these rules, but it does not increase the registered contribution-frequency outcome. This supports a behavioral response to the rule without establishing the proposed link from punishment to sustained cooperation.

For mean contribution size, excluding the money-maximization check,
all four mechanism rules move mean contribution in the direction predicted by
their literatures, with absolute effects between 0.93 and 1.45 credits. Thus, the findings differ by outcome: R1 does not establish a change in contribution frequency, whereas R3 reduces it despite increasing mean contribution size. Neither result should be summarized as no effect on cooperation overall. Matching runs by the number of completed periods shifts the effects by less than 0.8 percentage points and leaves the ranking unchanged. The leading rule is the same within each of the four admitted models; the conclusions for the other four rules vary by model.

\subsection{Stability across configurations}
\label{sec:pg-cost}

The full-sample analysis gives one set of effect estimates and rankings. We now ask how consistently smaller samples recover those conclusions, and whether the comparison stabilizes as more configurations are included.

\textbf{Method.}
For each sample size $n$, we draw 10,000 model-balanced resamples with replacement from the 268 configurations and repeat the analysis, keeping each rule run paired with its baseline. The full sample is a reference for agreement, not a known true ranking.

For each resample, we record the control-power estimates and three forms of agreement: whether it identifies the same leading rule, whether it preserves the order of the rule pairs distinguished by the full sample, and whether it reaches the same effect conclusion for every rule. An effect conclusion records whether the rule's signed 90\% interval lies entirely above zero, that is, whether the sample supports the registered prediction; an interval that crosses zero and one that lies entirely on the opposite side both count as not supported. We treat two rules as distinguished when the 90\% bootstrap interval for their difference in control power excludes zero. Among all five rules, the leader is the one with the highest control power. A resample of size 268 can repeat some configurations and omit others, so it need not reproduce the full-sample result exactly.

\begin{figure}[!htb]
  \centering
  \includegraphics[width=\linewidth]{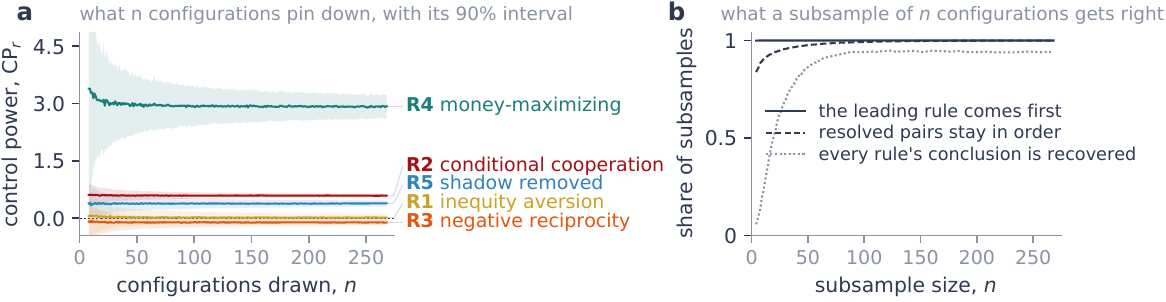}
  \caption{\textbf{How much the rule comparison changes with sample size.}
  We repeatedly draw $n$ configurations from the 268 available, with replacement
  and equal representation of the four admitted models.
  \textbf{(a)} Lines show mean control power and bands show its 90\% interval
  across resamples. Narrower bands indicate more precise estimates.
  \textbf{(b)} Three curves show agreement with the full-sample analysis:
  selecting the same leading rule, preserving the order of the ten rule pairs
  that the full sample distinguishes, and reaching the same effect conclusion
  for all five rules. An effect conclusion records whether its signed interval
  lies entirely above zero, in the registered direction. The full sample is a
  reference for stability, not a known true ranking.}
  \Description{Two panels side by side. The left panel plots the control power of
  five rules against the number of configurations drawn, with a band around each
  line and the rule named at the end of it; every line runs almost level from
  left to right while its interval narrows, one rule sits far above the rest and
  the other four lie close together. The right panel plots three shares against
  the number of configurations: the leading rule is recovered in essentially
  every resample, agreement across resolved rule pairs rises from about 84\% to
  nearly 100\%, and recovery of all five effect conclusions climbs from near
  zero, passes 80\% at forty configurations, and reaches 94\% at 268.}
  \label{fig:pg-convergence}
\end{figure}

\textbf{Results.}
The leading rule is recovered in 99.91\% of resamples at four
configurations and in every resample from eight configurations onward
(Figure~\ref{fig:pg-convergence}b). The full sample distinguishes all ten
rule pairs by control power; agreement across those pairs rises from 83.8\% at four
configurations to 99.98\% at 268. Conditional cooperation (R2) and removing the shadow of the future (R5) are closest, with control power of 0.59 and 0.38.

Recovering every rule's effect conclusion requires more evidence. All five
conclusions match the full sample in 51.9\% of resamples at twenty
configurations, 80.4\% at forty, 90.3\% at sixty-one, and 94.2\% at 268. The
remaining disagreement comes primarily from R1, whose full-sample effect is
near zero and whose resampled interval can therefore cross the decision
boundary. A study can thus identify the leading rule well before it can recover
the conclusion for every rule in the slate.

\subsection{Sensitivity to wording and position}
\label{sec:pg-wording}

The preceding estimates average over registered wordings at a fixed sentence position. We now ask how the effect depends on where the sentence appears in the profile and how many alternative wordings are included in the estimate.

\begin{figure}[!htb]
  \centering
  \includegraphics[width=\linewidth]{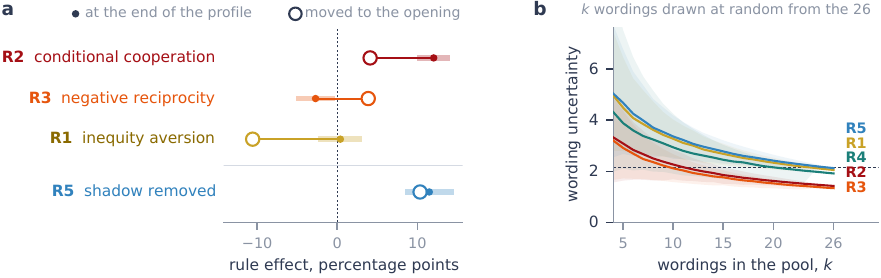}
  \caption{\textbf{How sentence position and wording affect the measured rule effect.} \textbf{(a)} Each rule's
  effect at the registered end of the participant profile (filled) and with the
  same sentence moved to just after its opening (hollow), in percentage points. The hollow dot is the registered effect plus the mean position shift estimated on the 28 matched configurations, not a separate estimate.
  The pale bar is the 90\% interval of the registered effect. Money-maximizing is
  not drawn: its effect is 78.6 points, and moving it changes the effect by only
  0.2 points. A larger distance between the filled and hollow dots indicates
  greater sensitivity to sentence position. \textbf{(b)} Estimated uncertainty
  due to the number of wordings used to measure each rule's effect
  (Section~\ref{sec:protocol}). It is estimated by drawing $k$ of
  the twenty-six at random: the line is the mean over draws and the band the
  middle 80\%. All five rules appear here. The dashed line is the target of a
  tenth of the baseline standard deviation; values below it meet the study's
  wording-uncertainty target.}
  \Description{Of the two panels, the left one shows one row per rule with a
  filled dot at the rule's registered effect, a hollow dot at its effect after
  the sentence was moved to the opening of the profile, and a line between them;
  inequity aversion's hollow dot lies well below zero, conditional cooperation's
  is a third of the way back toward zero, and the other two sit on their filled
  dots. The right panel plots five curves of wording uncertainty against the
  number of wordings in the pool, from four to twenty-six, each with a band;
  two curves fall below a dashed target line before the pool is half full and
  the other three reach it near the full pool.}
  \label{fig:pg-wording}
\end{figure}

\textbf{Method.}
To test position sensitivity, we move each rule sentence from the end of the participant profile to just after its opening, then compare its effect with that at the registered position (Figure~\ref{fig:pg-wording}a). We compare 28 matched configurations, each with one run per rule at each sentence position. The shift is the mean within-configuration difference between the two positions over these 28 configurations; we report it against each rule's full-sample effect at the registered position.

To assess wording uncertainty, we draw subsets of $k$ wordings from each rule's twenty-six registered wordings. For each subset, we calculate the uncertainty defined in Section~\ref{sec:protocol}: the standard deviation of the wording-level effects divided by $\sqrt{k}$. Figure~\ref{fig:pg-wording}b summarizes this quantity across draws. The target is one tenth of the baseline standard deviation. A wording's effect is estimated from the 5--22 configurations in which it appears. Because each wording appears in a different mix of models, we remove each model's mean effect and add back the rule's overall mean before comparing wordings, so each wording-level effect keeps the rule's level; the directions and ranges reported below use these adjusted estimates. Persona and world wordings are assigned at random and are not adjusted. For each value of $k$, we draw 2,000 subsets without replacement.

\textbf{Results.}
Three of the five effects change little when the sentence moves to the opening of the profile (Figure~\ref{fig:pg-wording}a). Conditional cooperation (R2) retains about a third of its effect, losing 7.9 percentage points from the original 12.0. Inequity aversion (R1) loses 10.9 points: its estimate changes from near zero at the end of the profile to a suppressive effect near the opening.

For wording uncertainty, every rule is under the target at the full pool, and the rules differ in how many wordings
that takes. Negative reciprocity (R3) and conditional cooperation (R2) are
under it with fewer than half the pool. Inequity aversion (R1) and the removed
shadow of the future (R5) need nearly all of it. R1 varies most across
wordings: half move the outcome in the opposite direction, and the individual
wording effects range from $-1.18$ to $+0.76$ baseline standard deviations.
Money maximization (R4) is much less sensitive: every wording moves the outcome
in the registered direction, and the weakest effect still exceeds two baseline
standard deviations.\\


\section{Explanatory support}
\label{sec:four}

Sections~\ref{sec:case} and~\ref{sec:pg} report behavioral records alongside outcome effects in their own worlds. To address RQ3, this section adds two interaction settings in which a single link of a proposed mechanism can be tested: group discussion, motivated by research on group polarization~\citep{moscovici1969group}, and cooperation among changing partners, motivated by research on cooperative cascades~\citep{fowler2010cooperative}. Both studies follow the calibration and paired comparisons in Section~\ref{sec:protocol}, testing five candidate rules per world, each with five registered wordings (Table~\ref{tab:four-slate}). Sections~\ref{sec:four-e2} and~\ref{sec:four-e3} present the setup and results for group polarization and cooperative cascades, respectively; Section~\ref{sec:four-close} brings together the findings across all four worlds.

\begin{table}[!htb]
  \centering
  \caption{Candidate rules for the two additional worlds. Each row shows one of five
  registered wordings. The selected wording is added to every participant's
  profile in the rule condition. Rules marked $\downarrow$ predict a decrease
  in the outcome; the other rules predict an increase.}
  \label{tab:four-slate}
  \small
  \renewcommand{\arraystretch}{1.14}
  \begin{tabular}{@{}
      >{\raggedright\arraybackslash}p{0.04\columnwidth}
      >{\raggedright\arraybackslash}p{0.31\columnwidth}
      >{\raggedright\arraybackslash}p{0.59\columnwidth}@{}}
    \toprule
    & Mechanism & Sample injected sentence \\
    \midrule
    \multicolumn{3}{@{}l}{\textbf{Group polarization}} \\[2pt]
    R1 & Value expression \rulesource{moscovici1969group}
       & ``You would rather say what you actually think than land somewhere in
         the middle.'' \\[2pt]
    R2 & Persuasive arguments \rulesource{burnstein1977persuasive,myers1976group}
       & ``You find it easier to agree with an argument you have not heard
         before.'' \\[2pt]
    R3 & Conformity pressure $\downarrow$ \rulesource{deutsch1955study}
       & ``You would rather not be the one holding the group up.'' \\[2pt]
    R4 & Involvement \rulesource{moscovici1969group}
       & ``You care about this question.'' \\[2pt]
    R5 & Fact-or-opinion framing \rulesource{moscovici1969group}
       & ``This is a question of fact, not of opinion.'' \\
    \midrule
    \multicolumn{3}{@{}l}{\textbf{Cooperative cascades}} \\[2pt]
    R1 & Behavioral contagion \rulesource{fowler2010cooperative}
       & ``You tend to do what the people around you were doing.'' \\[2pt]
    R2 & Descriptive norms \rulesource{cialdini1990focus}
       & ``You think of what people put in as what people here normally do.'' \\[2pt]
    R3 & Indirect reciprocity \rulesource{nowak2005evolution}
       & ``You expect to be treated the way you last saw people treated.'' \\[2pt]
    R4 & Social loafing $\downarrow$ \rulesource{Latane_1979,Karau_1993}
       & ``You find that one person easing off makes little difference to how
         the group does.'' \\[2pt]
    R5 & Threshold, raised $\downarrow$ \rulesource{granovetter1978threshold}
       & ``You wait for most of the others to put in before you do.'' \\
    \bottomrule
  \end{tabular}
  \Description{A table of ten candidate rules in two sections, one section
  per study. Each row gives a rule label, the mechanism it stands for with its
  literature source, and the exact sentence appended to every participant's
  profile. The two sections cover group polarization and cooperative cascades.}
\end{table}

\subsection{Group polarization}
\label{sec:four-e2}
\label{sec:four-results}

Moscovici and Zavalloni~\citep{moscovici1969group} asked participants to rate their opinions individually and then discuss them as a group to reach agreement. They compared the group's agreed position with members' initial ratings and collected individual ratings again after discussion. The collective pattern of interest is group polarization: discussion moves opinions further in the direction the group initially favored.

\textbf{Method.}
Our world has four participants whose initial ratings are 0, 1, 2, and 3 on a scale from $-3$ to $+3$, giving a mean of $+1.5$. They discuss one question until they agree. The outcome is the mean of members' post-discussion individual ratings minus the initial mean. A positive shift indicates movement toward the positive extreme; a negative shift indicates movement in the opposite direction. We also record speaking turns to distinguish changes in discussion activity from changes in opinion.

Calibration admits three models (Appendix~\ref{app:models}). Two operation settings were calibrated: the initial-rating spread and the discussion instruction. For the first, we compared three spreads, using twelve baseline runs per setting. In these baselines, discussion occurred under all three, but none polarized the group; at the widest spread, initial ratings of 0, 1, 2, and 3, consensus instead moved 0.58 scale points toward the center. We retained initial ratings of 0, 1, 2, and 3 because this setting left the most headroom for the outcome to move in either direction. For the second, we compared plain and step-by-step discussion instructions on two models; because the added steps did not materially change the discussion, we retained the simpler instruction. Twelve further baselines at the selected settings completed calibration. This baseline meets the calibration requirement: it need not polarize, but the shift can move in either direction. Further steps evaluate twenty-four configurations. Each of the three admitted models contributes eight configurations. Every configuration contains one baseline and one run for each rule. Four rules predict greater polarization: value expression, persuasive arguments, involvement, and fact-or-opinion framing. Conformity pressure predicts less. Value expression has twenty-three usable configurations because one rule run contains no final rating.

\textbf{Results.}
The baseline stays near the initial group position: the mean shift is $-0.09$ scale points across twenty-four configurations, with a standard deviation of 0.56 among them. Conformity pressure (R3), registered as suppressive, lowers the shift by 0.23 scale points and is the only established effect. None of the other rules establishes an increase in polarization. Persuasive arguments (R2) moves the shift by 0.19 points, while involvement (R4) moves it in the opposite direction to its prediction (Table~\ref{tab:four-ranking}).

The traces help characterize the gap between increased discussion and opinion change. In twenty-one of the twenty-four baseline runs, the final group mean falls between 1 and 2; 69\% of individual ratings also end in that range. This pattern is consistent with reaching agreement through compromise. Value expression (R1) increases discussion to two and a half times the baseline level while leaving ratings near the same middle. Baseline estimates also depend on wording: one profile wording gives a shift of $-0.44$, whereas the pooled estimate is $-0.09$. Averaging across the registered wording pool thus changes the starting point against which the candidate rules are assessed~\citep{clark1973language}.

The comparison therefore supports increased discussion under value expression, but not the predicted increase in polarization. A researcher transferring a discussion-based explanation to LLM agents should distinguish generating more discussion from changing how that discussion shapes opinions.

\subsection{Cooperative cascades}
\label{sec:four-e3}

Fowler and Christakis~\citep{fowler2010cooperative} studied whether cooperation spreads from person to person across successive interactions. They analyzed public-goods experiments in which participants contributed to a shared project in groups of four, observed their group's contributions, and then moved to new groups, with no pair meeting twice. The collective pattern of interest is a cooperative cascade: participants who observe greater contributions give more in later groups, spreading cooperation to people who never met the original contributor.

\textbf{Method.}
Our world has sixteen participants at four tables of four for five periods. Tables change each period according to a fixed schedule, and no two participants meet twice. Each table plays the public-goods game described in Section~\ref{sec:pg-world}, without punishment. Contributions are private during each decision. Between periods, a screen shows participants the contributions from the table they just left. The outcome is the mean number of credits contributed. We also examine first-period contributions, before participants have seen the screen, to distinguish initial responses to a rule from later responses to social information.

Calibration fixes the information channel and admits four models (Appendix~\ref{app:models}). Two operation settings were calibrated: the task description and the description of how participants received their credits. For the first, we compared three task descriptions using twelve baselines per description. In our runs, a money-maximizing instruction drove contributions near zero, whereas the neutral instruction left room for both increases and decreases, so we retained the latter. For the second, alternative descriptions produced little change, and we retained the simpler wording. The population of sixteen participants and five-period duration were fixed rather than calibrated. Further steps evaluate sixteen configurations. Each of the four admitted models contributes four configurations. Every configuration contains one baseline and one run for each rule. Behavioral contagion, descriptive norms, and indirect reciprocity predict increased contribution. Social loafing and the raised threshold predict decreased contribution. The table schedule and information channel remain fixed within every baseline--rule pair and across configurations.

\textbf{Results.}
The baseline contribution is 5.93 credits across four admitted models, with a standard deviation of 4.05 among the configurations, and it remains near that level over the five periods. Behavioral contagion (R1) raises contribution by 1.76 credits, descriptive norms (R2) raises it by 0.92, and social loafing (R4) lowers it by 1.80. All three effects are established. Indirect reciprocity (R3) and the raised threshold (R5) do not establish an effect.

All three supported effects are already present in the first period, before participants have seen the information screen (Table~\ref{tab:four-ranking}). These results support effects on contribution levels, but the first-period effects cannot arise from observing previous contributions. The overall increase therefore does not by itself establish that cooperation spreads across groups. Later social transmission remains possible; testing it requires a separate intervention on social information.

\begin{table}[!htb]
  \centering
  \caption{Rule effects in the two worlds, ranked within each world by control
  power. $\widehat{\Delta Y}_r$ is the rule effect in the metric's own units,
  signed by the registered direction ($\downarrow$ marks rules registered to
  lower the metric), and the next column reports whether the effect is
  established, that is, whether its signed 90\% interval lies entirely above zero. The last
  column is read from the traces: in
  the polarization world, speaking turns per run, against 2.5 in the baseline;
  in the cascade world, the contribution in the first period, before any screen
  has been shown, against 5.7 in the baseline. Value expression is read on
  twenty-three configurations because one of its runs leaves no rating.}
  \label{tab:four-ranking}
  \small
  \setlength{\tabcolsep}{4.5pt}
  \begin{tabular}{@{}llrrcr@{}}
    \toprule
    & Rule & $\widehat{\Delta Y}_r$ & $\mathrm{CP}_r$ & Established & In the traces \\
    \midrule
    \multicolumn{6}{@{}l}{\textbf{Group polarization} \ (signed shift, scale points; $N=24$; traces: speaking turns per run)} \\
    1 & R3 conformity pressure $\downarrow$ & $+0.23$ & $\mathbf{+0.39}$ & yes & 2.3 \\
    2 & R2 persuasive arguments    & $+0.19$ & $+0.30$ & no & 2.3 \\
    3 & R1 value expression        & $+0.09$ & $+0.13$ & no & 6.2 \\
    4 & R5 fact-or-opinion framing & $-0.03$ & $-0.04$ & no & 2.8 \\
    5 & R4 involvement             & $-0.08$ & $-0.15$ & no & 3.8 \\
    \multicolumn{6}{@{}l}{\textbf{Cooperative cascades} \ (contribution, credits; $N=16$; traces: first-period contribution)} \\
    1 & R4 social loafing $\downarrow$ & $+1.80$ & $\mathbf{+0.57}$ & yes & 4.5 \\
    2 & R1 behavioral contagion    & $+1.76$ & $\mathbf{+0.55}$ & yes & 7.3 \\
    3 & R2 descriptive norms       & $+0.92$ & $\mathbf{+0.42}$ & yes & 6.4 \\
    4 & R5 threshold raised $\downarrow$ & $+0.13$ & $+0.05$ & no & 5.6 \\
    5 & R3 indirect reciprocity    & $-0.43$ & $-0.18$ & no & 5.1 \\
    \bottomrule
  \end{tabular}
  \Description{A table of rule effects in two sections, one per study, each
  ranked by control power. The polarization section is read on the signed shift
  and the cascade section on contribution in credits. Each row gives the rule
  effect, its control power, whether the effect is established, and one number
  read from the traces: speaking turns per run for the polarization rules and
  first-period contribution for the cascade rules.}
\end{table}

\subsection{Findings across the four worlds}
\label{sec:four-close}
\begin{figure}[!htbp]
\centering
\includegraphics[width=\linewidth]{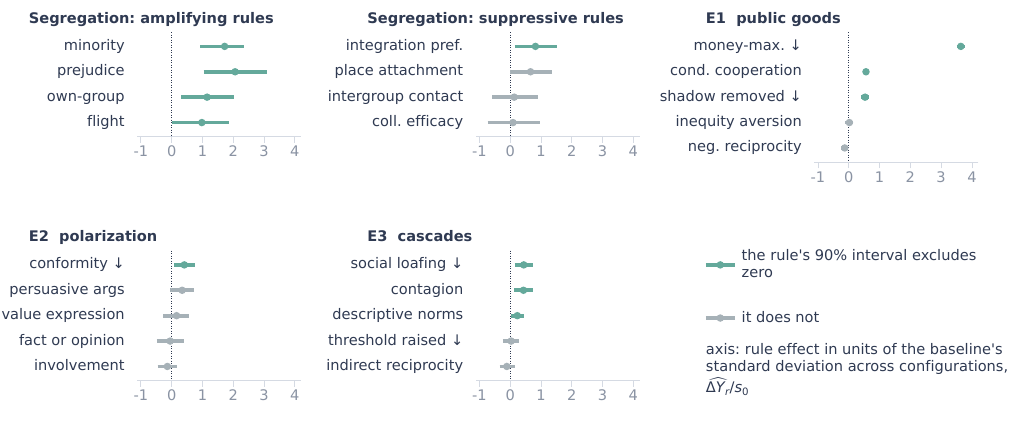}
\caption{\textbf{Which rules change collective outcomes across the four worlds?}
Each dot shows a rule's mean change from its paired baseline. To place different
outcomes on one scale, this change is divided by that world's baseline standard
deviation across configurations, $s_0$, giving Glass's $\Delta$. Positive values mean change in the
registered direction, including a decrease for suppressive rules. Bars show
90\% intervals: dark bars lie entirely in the registered direction, while light
bars do not support an effect in that direction. Place attachment's interval reaches zero. Rules are ordered by
control power within each panel; the segregation rules are shown separately
for amplifying and suppressive directions. The horizontal scale uses baseline-standardized effects, whereas the ordering uses control power, standardized by the variation in paired differences. These comparisons concern outcome shifts, not evidence that the proposed social process occurred.}
\Description{The figure is a forest plot with five panels. The first two panels
show the segregation rules on the departure-rate gap: the four amplifying
rules, whose intervals all lie in the registered direction, and the four suppressive rules, of
which only integration preference does. The third panel shows the five public-goods rules, three
of whose intervals lie in the registered direction. The last two panels show the polarization and cascade
worlds, where one interval lies in the registered direction in the polarization world and three in
the cascade world.}
\label{fig:forest}
\end{figure}

Each of the four worlds yields at least one rule with an established effect in its registered direction, and three yield such effects in both directions. Across all twenty-three candidate rules, twelve have 90\% intervals entirely in their registered direction (Figure~\ref{fig:forest}).

The method identifies rules that change measured outcomes across all four interaction settings. These effects do not by themselves establish that a simulation reproduces the source phenomenon: more discussion does not necessarily produce polarization, and higher contributions do not establish a cooperative cascade. The behavioral records therefore qualify the explanatory interpretation of the outcome effects. In the polarization world, increased discussion does not produce more extreme opinions: agents interact more, but the collective outcome does not follow. In the cascade world, contribution effects precede exposure to others' behavior: the outcome changes before interaction could have caused it. The method distinguishes evidence that a rule changes an outcome from evidence about the behavioral links proposed to explain that change. For researchers adopting LLM agents, the implication is to state which link has been tested: a response to a rule, a change in the collective metric, or the interaction connecting them. Unresolved links identify targets for follow-up interventions, not grounds for either accepting or rejecting the entire social theory. Each finding applies to the baseline world and configuration pool tested.

\section{Discussion and Guidelines}
\label{sec:discussion}

Across four simulation worlds, we find that natural-language rules can produce measurable collective effects, but that the strength and interpretation of those effects depend on how the simulation is constructed and evaluated. Some conclusions stabilize across configurations, while others change with rule wording or sentence position. Behavioral records further show that an outcome effect can occur without the social process proposed to explain it. These findings lead to four guidelines for using generative agent-based models to explore causal mechanisms.\\
\textbf{Calibrate for the research question, not the expected pattern.}
Specify the actions, interactions, and outcome range required for the research question before testing rules. In our segregation study, unrestricted action budgets yielded few relocation decisions; calibration made the relevant behavior available for comparison. In our polarization study, valid discussions did not yield a more extreme group position, but the outcome still left room to test rule effects. These cases require different responses: execution faults call for repair, whereas an unexpected baseline need not be tuned until it resembles the source phenomenon. Researchers testing whether a rule produces a pattern may begin from its absence; those asking whether an existing pattern can be strengthened or weakened need a baseline that exhibits it. Fixing the calibrated world before rule testing prevents outcome-driven changes to the experimental setup.\\
\textbf{Treat rule wording as part of the experimental design.}
Distinguish the proposed rule from the sentence used to express it. In our public-goods study, some wordings of inequity aversion produced effects in opposite directions, and moving the sentence changed the estimated effect. Repeating a successful sentence cannot establish that its effect generalizes across expressions. Researchers should specify the wording pool and sentence position, check that variants preserve the intended claim, and report variation alongside the pooled estimate. Matching models and background prompts controls those differences within a comparison; it does not remove sensitivity to how the rule itself is expressed. Claims should remain bounded by the evaluated models and expressions.\\
\textbf{Match the evidence budget to the intended conclusion.}
Decide whether the study needs to estimate an average effect, select a leading rule, or distinguish all candidates. These tasks can require different amounts of evidence: in our public-goods study, the leading rule stabilized before every rule's effect conclusion did. Report effect magnitude, variation across configurations, and estimation uncertainty separately. A larger sample can make an inconsistent effect more precisely estimated without making it more consistent. Resampling can assess agreement with the available reference sample, but does not establish a universally sufficient sample size or a true ranking. Where rules remain difficult to distinguish, report their uncertain order.\\
\textbf{Test the link from behavior to outcome.}
State the behavioral sequence proposed to explain the collective result, and examine which links the records support. In our public-goods study, negative reciprocity increased punishment and mean contribution size without increasing contribution frequency; these measures support different claims about cooperation. In our other studies, increased discussion did not yield polarization, and contribution effects in the cascade world preceded exposure to social information. Researchers should therefore examine behavior and timing alongside the registered outcome, without replacing that outcome after seeing results. Where a link remains unresolved, design a follow-up intervention on it---for example, varying access to others' contributions to test later social transmission. Keep exploratory observations distinct from registered tests. A compatible trace strengthens the interpretation of a rule effect, but does not by itself establish that the proposed process is necessary or uniquely responsible.

\section{Scope and Ethical Considerations}
\label{sec:scope-ethics}
Our evidence concerns interventions within generative agent-based models. We estimate how collective outcomes change when a candidate mechanism is operationalized as a natural-language rule and introduced into a calibrated simulation world. Agents may represent individuals, households, organizations, or other entities. The interpretation of a result therefore depends on what the agents represent and how the simulation world maps onto the system of interest. The simulations do not establish that agents behave like their real-world counterparts. The four studies examine individual mechanisms under fixed interaction structures; combinations of mechanisms, alternative population structures, and longer-term dynamics remain open for future work.

Generative ABMs may reproduce assumptions and biases encoded in their models, prompts, and world designs~\citep{wang2025flatten}. This risk is especially important when agents represent people or social groups, as in our residential segregation study. Simulation results can help researchers compare and refine hypotheses, but they should not be treated as direct evidence about the represented entities~\citep{agnew2024illusion}. 


\section{Conclusion}
\label{sec:conclusion}

Generative agent-based models can do more than produce collective patterns: they can help researchers compare candidate mechanisms that could produce those patterns within a simulation. RePair calibrates and fixes a simulation world, operationalizes candidate mechanisms as natural-language rules, estimates their effects through matched comparisons, and examines behavioral evidence for the proposed causal process. Across four worlds, we show that the method can distinguish candidate mechanisms by their collective effects, assess how reliably those comparisons hold across implementations, and reveal when an outcome effect does not support the proposed causal process. It provides a structured way to screen, compare, and refine micro-to-macro explanations.

\bibliographystyle{IEEEtran}
\bibliography{references}
\appendix
%
%

\section{The models considered}
\label{app:models}

Each world screens the models declared for it and admits those whose baseline
executes validly and leaves the outcome room to move in both directions
(Section~\ref{sec:phase1}). In the segregation world the screen had two stages.
Each candidate model first ran one baseline, which had to pass the execution
checks. It then ran four more, whose pooled departures had to average about
seven per baseline, the calibration threshold. A model that could not complete a verified baseline
was excluded. The screen is per world, so a model admitted in one world may be excluded in another. The segregation world admitted DeepSeek V4 Flash, GLM-4.7 Flash, Qwen 3.7 Flash, and Qwen 3.7 Plus from twelve screened models. The public-goods world admitted DeepSeek V4 Flash, GLM-4.7 Flash, Gemma 4 26B, and Gemma 4 31B from seven. The polarization world admitted DeepSeek V4 Flash, Qwen 3.7 Flash, and Gemma 4 26B from seven. The cascade world admitted DeepSeek V4 Flash, GLM-4.7 Flash, Qwen 3.7 Plus, and Gemma 4 26B from seven. Every result reported in this paper rests on the admitted models.

\end{document}